\documentstyle[preprint,floats,epsfig,aps]{revtex}
\begin{document}
\def\DESepsf(#1 width #2){\epsfxsize=#2 \epsfbox{#1}}
\input{epsf}

\begin{center}
 \vskip 15mm
{\large Neutrino Masses with ``Zero Sum'' Condition: 
$m_{\nu_1} + m_{\nu_2} + m_{\nu_3} = 0$ }
 \vskip 15mm
Xiao-Gang He$^a$ and A. Zee$^b$\\
$^a$ Department of Physics, National Taiwan University, Taipei, Taiwan\\
$^b$ Institute for Theoretical Physics, University of California,
Santa Barbara, California 93106, USA
\vskip 15mm
\end{center}

\begin{abstract}
It is well known that the 
neutrino mass matrix contains more parameters than experimentalists 
can hope to 
measure in the foreseeable future even if we impose CP invariance. 
Thus, various authors have proposed ansatzes to restrict the 
form of the neutrino mass matrix further. Here we propose that
$m_{\nu_1} + m_{\nu_2} + m_{\nu_3} = 0$; this ``zero sum'' condition can
occur in certain class of models, such as models
whose neutrino mass matrix can be expressed as commutator
of two matrices.
With this condition, the absolute neutrino mass can be
obtained in terms of the
mass-squared differences. When combined with the accumulated
experimental data this condition predicts two types of
mass hierarchies,
with one of them
characterized by
$m_{\nu_3} \approx -2m_{\nu_1} \approx -2 m_{\nu_2}
\approx 0.063$ eV,
and the other by
$m_{\nu_1} \approx -m_{\nu_2} \approx 0.054$ eV and
$m_{\nu_3} \approx 0.0064$ eV.
The mass ranges predicted is just below the cosmological 
upper bound of 0.23 eV from recent
WMAP data and can be probed in the near future. 
We also point out some implications for
direct laboratory measurement of neutrino masses, and the neutrino mass
matrix.
\end{abstract}

\newpage

\section{Introduction}

There are abundant data\cite{1,2,3,4,5,6} from solar,
atmospheric, laboratory and long baseline neutrino
experiments on
neutrino mass and mixing.
The present experimental data, including recent
results from
KamLAND\cite{5} and K2K\cite{6}, on neutrino masses and mixing angles
can be
summarized as follow\cite{8,9,10}.
The 99.3\% C.L. level allowed ranges for the mass-squared
differences are constrained to be: $1.5\times 10^{-3}$
eV$^2$
$\leq \Delta m^2_{atm} \leq 5.0\times 10^{-3}$ eV$^2$,
and
$2.2\times 10^{-5}$ eV$^2$ $\leq \Delta m^2_{solar} \leq
2.0 \times 10^{-4}$ eV$^2$, with the best fit values
given by
$\Delta m^2_{atm} = 3.0\times 10^{-3}$ eV$^2$, and
$\Delta m^2_{solar}
= 7.0\times 10^{-5}$ eV$^2$.
The mixing angles are in the ranges of $\sin^22\theta_{atm}
> 0.85$
and $0.18 \leq \sin^2\theta_{solar} \leq 0.37$. Also
the CHOOZ
experiment\cite{4} gives a bound of $0.22$ on the $\nu_e -\nu_x$ 
(where $\nu_x$ can be either $\nu_\mu$ or $\nu_\tau$ or a 
linear combination) oscillation parameter. 

Present data can be explained by oscillations between three
active neutrinos\cite{osc} \footnote{There are additional evidences for 
oscillation between electron and muon neutrinos 
from LSND experiment\cite{11}. If confirmed more 
neutrinos are needed to explain all the data.}
with the atmospheric neutrino and K2K data explained by
oscillation between the muon
and the tauon neutrinos, and the solar neutrino and KamLAND
data explained
by oscillation between the electron and muon neutrinos.
Neutrino oscillations provide direct evidence of non-zero neutrino
masses and
mixing between different species of neutrinos. 

We will assume that the neutrinos are Majorana as favored by some 
theoretical considerations \cite{7}. 
The relevant term in the Lagrangian
describing Majorana neutrino masses 
is $\cal {L}$ $= 
\nu_L^T C M \nu_L + H.C.$, where $\nu_L = (\nu_{eL}, \nu_{\mu L}, 
\nu_{\tau L})$ are the left-handed 
neutrinos and $C$ is the charge conjugation operator. 
The mass matrix $M$ is symmetric due to Fermi statistics. 
A convenient basis to study
neutrino masses and mixing is the weak basis where all
charged leptons are already diagonalized. The unitary
mixing matrix $V$ and the mass eigenvalues in this basis
are related by

\begin{eqnarray}
D = V^T M V,
\label{m}
\end{eqnarray}
where $D$ is a diagonal matrix. The diagonal entries $m_i$ of $D$ 
are the mass
eigenvalues which can always be made real by an appropriated
choice of phase
convention.  For three generations of neutrinos, the
mixing matrix
can be described by three rotation angles and three phases.
Note that in general $TrM \neq Tr D$.

Although there is a lot of data on neutrinos, more data are
needed to determine detailed
properties of neutrinos. Oscillation experiments, as
indicated above, cannot measure the values $m_{i}$, but only
the mass-squared differences
$\Delta m^2_{21} = m_{\nu_2}^2 - m_{\nu_1}^2$ and $\Delta m^2_{32} = 
m_{\nu_3}^2 - m_{\nu_1}^2$ and some mixing angles.
In analyzing the present data, it is generally assumed that
$\Delta m^2_{21}
\equiv \Delta m^2_{solar}$, and $\Delta m^2_{32} \equiv
\Delta m^2_{atm}$ and
thus $|\Delta m^2_{21}| \ll |\Delta m^2_{32}|$. The sign of 
$\Delta m^2_{21}$ has now been determined to be positive from matter 
effect in the solar
neutrino data\cite{2,sign}, but the sign of $\Delta m^2_{32}$ is not
determined yet.

There is at present certainly no information on any of the
three CP violating phases, and in the foreseeable future
no set of experiments can fully determine all the parameters
in the neutrino mass matrix.
Certain theoretical inputs have to be employed to reconstruct
the neutrino
mass matrix\cite{12,13,14,15,16,17,18}. 
Several proposals have been made to reduce the
parameters, such as texture zero\cite{14} and determinant zero
requirement\cite{15} for the mass matrix.

Here we propose another way of reducing the number of unknown parameters,
by imposing a condition on the mass eigenvalues,

\begin{eqnarray}
m_{\nu_1} +m_{\nu_2} +m_{\nu_3}=0.
\label{cond}
\end{eqnarray}
This ``zero sum'' condition that the neutrino masses sum to zero 
implies that some of the
eigen-masses must have opposite signs. These signs of course
can always be arranged
by making chiral phase redefinition of the neutrino fields.

If there is no CP violation in the neutrino mass matrix $M$, the
mass matrix can always be made real and it can be diagonalized
by an orthogonal transformation, namely $V^T V = I$. In this case the
``zero sum'' condition (\ref{cond}) 
is equivalent to the traceless condition,
$TrM = Tr(V^\dagger V^* D) = Tr D =0$. If CP is not conserved, the ``zero sum''
and traceless conditions are different. One needs to be careful 
about the phase definitions\cite{nieves}.  
In this paper we simply explore the phenomenological
consequences of requiring the neutrino masses to satisfy the ``zero 
sum'' condition with CP conservation
without speculating on its theoretical origin.
We note, however, that it holds if $M = [A,B]$,
that is, the mass matrix can be expressed as a commutator\footnote{In 
the simplest
versions of models proposed in ref.\cite{12}, $M$ results from
radiative correction and comes out to be the commutator
of a coupling matrix and the mass-squared matrix of the
charged leptons. However there are a number of variations
of this model and the condition $m_{\nu_1} + m_{\nu_2} + m_{\nu_3} = 0$
fails to hold in most of them.} of two matrices $A$ and $B$. 

\section{``Zero Sum'' Condition and Neutrino Masses}

Direct measurement of neutrino masses is a very difficult
experimental
task. If the ``zero sum'' 
condition $m_{\nu_1} + m_{\nu_2} + m_{\nu_3} =0$
or equivalently $TrM =0$, is applied, all
the neutrino masses are determined in terms of the
mass-squared differences.
We have

\begin{eqnarray}
&&m_{\nu_1}^2 = -{1\over 3}\left [2 \Delta m^2_{21} +  \Delta m^2_{32}
\pm 2\sqrt{(\Delta m^2_{32})^2 + \Delta m^2_{21}\Delta m^2_{32} +
(\Delta m^2_{21})^2}
\right ],
\nonumber\\
&&m_{\nu_2}^2 = {1\over 3}\left [\Delta m^2_{21} - \Delta m^2_{32}
\pm 2\sqrt{(\Delta m^2_{32})^2 + \Delta m^2_{21}\Delta m^2_{32} +
(\Delta m^2_{21})^2}
\right ],
\nonumber\\
&&m_{\nu_3}^2 = {1\over 3}\left [\Delta m^2_{21} + 2\Delta m^2_{32}
\pm 2\sqrt{(\Delta m^2_{32})^2 + \Delta m^2_{21} \Delta m^2_{32} +
(\Delta m^2_{21})^2}
\right ].
\end{eqnarray}
The choice of the signs in front of the square root
is decided by the
requirement that all $m^2_{\nu_i}$ must be larger or equal to zero.
From the expression for $m^2_{\nu_3}$, one can determine that
the ``+'' has to be chosen for the expressions for $m_{\nu_{2,3}}$, and 
``$-$'' for the expression for $m_{\nu_1}$. The relative
signs of the eigen-masses $m_i$ are determined by the
condition (\ref{cond}). We will use a convention such that $m_{\nu_3}\geq 0$
in our later discussions.

For  $|\Delta m^2_{21}| \ll |\Delta m^2_{32}|$, the above
simplify to, to leading
order in $\varepsilon \equiv \Delta m^2_{21}/\Delta m^2_{32}$,
we have

\begin{eqnarray}
&&m_{\nu_1}^2 = {2\over 3} [|\Delta m^2_{32}| ( 1+ {\varepsilon \over 2})
- {1\over 2} \Delta m^2_{32}(1+ 2\varepsilon)],\nonumber\\
&&m_{\nu_2}^2 = {2\over 3} [|\Delta m^2_{32}| ( 1+ {\varepsilon \over 2})
- {1\over 2} \Delta m^2_{32}(1 - \varepsilon)],\nonumber\\
&&m_{\nu_3}^2 = {2\over 3} (|\Delta m^2_{32}|+ \Delta m^2_{32}) 
(1+ {\varepsilon\over 2}).
\end{eqnarray}

At present the sign of the measured $\Delta m_{21}^2$ is determined to be 
positive, but the sign of $\Delta m^2_{32}$ is not determined,
there are two possible solutions corresponding to the sign of 
$\Delta m^2_{32}$. In Table I we list
all solutions for the best fit values of the mass-squared
differences.

\begin{table}
\caption{ Solutions of eigen-masses for the best fit values of
$|\Delta m_{21}^2| = 7.0\times 10^{-5}$ eV$^2$ and $|\Delta m^2_{32}|
= 3.0\times 10^{-3}$ eV$^2$.}
\begin{center}
\begin{tabular}{|c|c|c|c|c|c|}
$\Delta m_{32}^2$ (eV$^2$)&$\Delta m^2_{21}$ (eV$^2$)&$m_{\nu_1}$ (eV)&
$m_{\nu_2}$ (eV)&
$m_{\nu_3}$ (eV)&$|m_{ee}|$ (eV)\\ \hline
$3.0\times 10^{-3}$& $7.0\times 10^{-5}$&$-0.0313$&$-0.0324$&0.0636&
$(0.01\sim 0.032)$\\
$-3.0\times 10^{-3}$& $7.0\times 10^{-5}$&0.0541&$-0.0548$&
$6.43\times 10^{-4}$&$(0.018\sim 0.054)$
\end{tabular}
\end{center}
\end{table}

We see that the mass eigenvalues exhibit two types of hierarchies,

\begin{eqnarray}
&&I)\;\;\;m_{\nu_3} \approx -2 m_{\nu_1} \approx -2 m_{\nu_2} 
\approx 0.064\;\;
\mbox{eV}\nonumber\\
&&II)\;\;\; m_{\nu_1} \approx - m_{\nu_2} \approx 0.054\;\;\mbox{eV},
\;\;\mbox{and}
\;\;m_{\nu_3} \approx 0.00064\;\;\mbox{eV}.
\label{hierarchy}
\end{eqnarray}
The sign of $\Delta m^2_{32}$ decides which mass hierarchy the
solutions belong to. Note that the ``natural'' sign $\Delta m^2_{32} >0$
corresponds to scenario I), in which the masses are of the same order
of magnitude, in contrast to scenario II), in which $m_{\nu_3}$ is two 
order of magnitude smaller than $m_{\nu_1}$ and $m_{\nu_2}$. We would like
to suggest that I) is more favored than II).

In contrast to oscillation experiments, the contribution of the
neutrinos to
the energy density of the universe, $\Omega_\nu
\approx \sum_i |m_i|/(90\;\mbox{eV}) $ depends on the absolute
values of $|m_i|$
of course. With the neutrino masses predicted with the central
values of
$\Delta m^2_{21}$ and
$\Delta m^2_{32}$, we would have, $\Omega_\nu \ll 1 $.
Even if one uses the upper bound of
$\Delta m^2_{32} = 5.0\times 10^{-3}$,
the contribution to the energy density is
still much smaller than one. Other astrophysical and
cosmological data can give information on the neutrino masses\cite{weiler}, 
such as the CMB power spectrum and large scale structure survey data.
The absolute neutrino mass sum
$|m_{\nu_1}| + |m_{\nu_2}| + |m_{\nu_3}|$ predicted by the ``zero sum''
condition is only a factor of two 
smaller than the present bound\cite{map} 
of 0.23 eV obtained from combining WMAP and 
Galaxy sky survey data, and is close to
the sensitivity of 0.12 eV of the combined
PLANCK CMB data with the SDSS sky survey\cite{planck}. A future
sky survey with an order of magnitude
larger survey volume would allow the sensitivity to reach
0.03 eV\cite{weiler}. The mass ranges predicted by the
condition (\ref{cond}) may be tested in the future.

\section{Mixing and Masses}

To obtain more information about neutrino properties,
one needs to
have information from mixing.
Before the SNO and KamLAND data, assuming three active
neutrino oscillations,
one of the solutions which can account for known data is 
the bi-maximal mixing matrix\cite{bi} with $|V_{e2}| = |V_{\mu 3}|
= 1/\sqrt{2}$.
Experimental data from SNO and the recent data from KamLAND 
however disfavor the maximal
mixing for the $V_{e2}$ entry.
We were thus led to propose\cite{19} the following mixing matrix,

\begin{eqnarray}
V = \left ( \begin{array}{rrr}
{-2\over \sqrt{6}}& {1\over \sqrt{3}}&0\\
{1\over \sqrt{6}}& {1\over \sqrt{3} }& {1\over \sqrt{2}}\\
{1\over \sqrt{6}} & {1\over \sqrt{3}}& -{1\over \sqrt{2}}
\end{array}
\right ).
\label{mixing}
\end{eqnarray}
This mixing matrix (but with the first and second column interchanged) 
was first suggested by Wolfenstein more than 20 years ago\cite{20}. 
It has subsequently been studied extensively by 
Harrison, Perkins and Scott\cite{21}, and Xing\cite{22}.

As mentioned before, oscillation experiments can not
determine the relative signs of the mass eigenvalues which
implies that one can multiply a phase matrix $P =
Diag(e^{i\rho}, e^{i\sigma}, 1)$ from the right on $V$.
With CP invariance, $\sigma$ and $\rho$ can take the values of zero or
$\pm \pi/2$.

With the above information on the mixing matrix,
let us estimate two observables
related to neutrino mass measurements, the effective
mass electron neutrino
mass $\langle m_e \rangle ^2$ measured in
tritium beta decay end point spectrum experiments,
and the effective Majorana electron neutrino mass $m_{ee}$
in neutrinoless double beta decays.

The effective mass measured in tritium end point spectrum
experiments is given by, $\langle m_e\rangle ^2 =
(m_{\nu_1}^2 |V_{e1}|^2 + m_{\nu_2}^2 |V_{e2}|^2 +
m_{\nu_3}^2 |V_{e3}|^2)$. Using the values for the neutrino
masses in Table I, we
find that $\langle m_e \rangle $ is below the
sensitivity of $0.12$ eV for the proposed
experiment KATRIN\cite{23}. However, neutrinoless double decay
experiments may
be sensitive to the predicted ranges.
The amplitude of neutrinoless double beta decay is
proportional to the effective electron-neutrino
Majorana mass $m_{ee}$, namely $M_{11}$, given by

\begin{eqnarray}
|m_{ee}| = |m_{\nu_1} V_{e1}^{*2} e^{-2i\rho} +
m_{\nu_2} V_{e2}^{*2} e^{-2i\sigma} + m_{\nu_3} V_{e3}^{*2}|.
\label{double}
\end{eqnarray}

From the above expression we see that the value $m_{ee}$ depends 
on the Majorana
phases $\rho$ and $\sigma$.
If the phases $\rho$ and $\sigma$ are
all zero, the neutrinoless double beta decays would have
the smallest
rate because the cancellation imposed by the ``zero sum'' 
condition. If the phases take values such that both 
$m_{\nu_1} e^{-2i\rho}$ and $m_{\nu_2} e^{-2i\sigma}$ are positive,
the neutrinoless double beta decays would have the
largest rate possible. To have an idea of what possible
effects of
the Majorana phases can have on neutrinoless double beta decays,
we calculated the range for $|m_{ee}|$ for the best fit
values of the
mass squared differences with $V$ given in eq.(\ref{mixing}) 
for arbitrary $\rho$
and $\sigma$.
The allowed ranges are also listed in Table I at the
last column.
The ranges obtained are well below the present
upper bounds
of 0.4 eV\cite{24}, but
can almost be fully covered by future experiments\cite{25}, such as
GENIUS, MAJORANA, EXO, MOON or COURE, where sensitivity
as low as
0.01 eV seems possible.

\section{Neutrino Mass Matrix}

We now study implications of the ``zero sum'' neutrino mass 
condition on the
neutrino mass matrix $M$. If the neutrino masses and mixing matrix are 
known to a good precision,
one can invert the eigen-masses according to eq. (\ref{m}) to
obtain the
mass matrix $M$. To have some feeling how this may provide important
information
about the mass matrix, we present some details of the mass
matrices which
produce the mixing matrix $V$ in eq. (\ref{mixing}).

The most general mass matrix which can produce the mixing
matrix $V$ in eq. (\ref{mixing})
can be specified in the following form by the mass eigenvalues,

\begin{eqnarray}
M = {m_{\nu_1} \over 6}
\left ( \begin{array}{ccc}
4& -2& -2\\
-2& 1& 1\\
-2& 1& 1
\end{array}
\right )
+
{m_{\nu_2} \over 3} \left ( \begin{array}{ccc}
1& 1& 1\\
1& 1& 1\\
1& 1& 1
\end{array}
\right )
+ {m_{\nu_3}\over 2}
\left ( \begin{array}{ccc}
0& 0& 0\\
0& 1& -1\\
0& -1& 1
\end{array}
\right ).
\label{mass}
\end{eqnarray}
Being real symmetric (and so a fortiori Hermitian)
the above three matrices generate a $U(1)\otimes U(1)\otimes U(1)$ 
subgroup of $U(3)$.  

With the ``zero sum'' condition $m_{\nu_1} + m_{\nu_2} + m_{\nu_3}=0$,
the above matrix can be written as

\begin{eqnarray}
M = {m_{\nu_1} \over 3}
\left ( \begin{array}{ccc}
2& -1& -1\\
-1& -1& 2\\
-1& 2& -1
\end{array}
\right )
+
{m_{\nu_2} \over 3} \left ( \begin{array}{ccc}
1& 1& 1\\
1& -1/2& 5/2\\
1& 5/2& -1/2
\end{array}
\right ).
\end{eqnarray}

Since $|\Delta m^2_{21}/\Delta m^2_{32}| \ll 1$,  it is
instructive to work
with the case $\Delta m^2_{21} = 0$ as the first approximation and
to see how
the obtained mass matrix can be perturbed to produce the
desired $\Delta m_{21}^2$.
With this approximation
the two mass hierarchies I) and II) are then
$(m_{\nu_1} = m_{\nu_2} = -m_{\nu_3/}2 = 2a)$, and
$( m_{\nu_1} = -m_{\nu_2} = 2a, m_{\nu_3} = 0)$, respectively.
The corresponding mass matrices are
given by

\begin{eqnarray}
I)\;\;\;\; M_0 = a
\left ( \begin{array}{ccc}
2& 0& 0\\
0& -1& 3\\
0& 3& -1
\end{array}
\right );\;\;\;\;
II)\;\;\;\;
M_0 = {1 \over 3} a\left ( \begin{array}{ccc}
2& -4& -4\\
-4& -1& -1\\
-4& -1& -1
\end{array}
\right ).
\label{m0}
\end{eqnarray}
Note that the unperturbed mass matrix $M_0$ looks ``simpler'' 
in the ``natural'' hierarchy I) than in the ``inverted'' hierarchy II).

The mass matrix in case I) has been studied in a previous paper
by us\cite{19}. The
desired mass-squared difference $\Delta m^2_{21}$ can be
obtained by a small
perturbation of the form

\begin{eqnarray}
\delta M_T = \varepsilon a
\left ( \begin{array}{ccc}
0& 1& 1\\
1& 0& 1\\
1& 1& 0
\end{array}
\right ),
\end{eqnarray}
with the perturbed eigenvalues given by
$m_{\nu_1} = 2a(1-\varepsilon/2)$, $m_{\nu_2} = 2a(1+\varepsilon)$, and
$m_{\nu_3} = -4a(1+\varepsilon/4)$. The parameter $\varepsilon$ to
the lowest order
is given by $\varepsilon = \Delta m^2_{21}/\Delta m^2_{32}$.

For case II), adding $\delta M_T$ to $M_0$, the eigen-masses
are given by
$m_{\nu_1} = 2a(1-\varepsilon/2)$, $m_{\nu_2} = - 2a(1-\varepsilon)$,
and
$m_{\nu_3} = -a \varepsilon$ with $\varepsilon \approx \Delta m^2_{21}/
\Delta m^2_{32}$.

The perturbation $\delta M_T$
preserves  the
``zero sum'' condition in eq. (\ref{cond}).
One can also consider situations where the perturbations
break this ``zero sum'' condition, but still produce the mixing
matrix $V$ in eq. (\ref{mixing}).
For example, adding a ``democratic" perturbation,

\begin{eqnarray}
\delta M_D = \varepsilon a
\left ( \begin{array}{ccc}
1& 1& 1\\
1& 1& 1\\
1& 1& 1
\end{array}
\right ),
\end{eqnarray}
produces a mixing matrix of the form given by $V$ in eq. (\ref{mixing}),
but different
mass eigenvalues,
($m_{\nu_1} = 2a$, $m_{\nu_2} = 2a(1+3\varepsilon/2)$, 
$m_{\nu_3} = -4a$,
$\varepsilon = \Delta m^2_{21}/\Delta m_{32}^2$), and
($m_{\nu_1} = 2 a$, $m_{\nu_2} = - 2a(1-3\varepsilon/2)$, 
$m_{\nu_3} = 0$,
$\varepsilon = \Delta m^2_{21}/3\Delta m^2_{32}$)
for case I) and case II), respectively.

We would like to comment that the reconstruction of
mass matrix depends crucially on knowing the mixing matrix 
even if all the eigenvalues for
the masses are known through experimental measurements
or theoretical
considerations such as $m_{\nu_1}+m_{\nu_2} +m_{\nu_3}=0$ or the traceless
condition for real Majorana neutrino mass matrix proposed here.
Had the mixing
matrix been given by the bi-maximal mixing matrix\cite{bi} which
was allowed by
pre-SNO and KamLAND data,
the mass matrix would be very different.
We mention that the simplest version of the model proposed in 
ref.\cite{12} with all diagonal
entries zero
in the mass matrix can easily produce a bi-maximal mixing,
but it is difficult
to produce the mixing matrix given by $V$ in eq. (\ref{mixing}).

\section{Conclusions}

We have studied the consequences of neutrino masses with
the ``zero sum'' condition
$m_{\nu_1} + m_{\nu_2} + m_{\nu_3} = 0$.
With this condition the neutrino masses can be determined
from measured
mass-squared differences from oscillation experiments.
We find that this condition predicts only two types of neutrino
mass hierarchies with one of them characterized by
$m_{\nu_3} \approx -2m_{\nu_1} \approx -2 m_{\nu_2} \approx 0.063$ eV,
and another by $m_{\nu_1} \approx -m_{\nu_2} \approx 0.054$ eV and
$m_{\nu_3} \approx 0.0064$ eV.
These masses although small, can be probed by experiments from
CMB measurements and
large scale structure survey, and can also be probed by
neutrinoless
double beta decay experiments. In conjunction with information
on neutrino
mixing, the ``zero sum'' condition also predicts simple mass
matrices for neutrinos.

\acknowledgments
We thank H. Murayama, S. Pakvasa, R. Volkas and L. Wolfenstein for 
helpful discussions and comments on the manuscript.
This work was supported in part by
NSC under grant number NSC 91-2112-M-002-42,
and by the MOE Academic Excellence Project 89-N-FA01-1-4-3 of Taiwan,
and by NSF under grant number PHY 99-07949 of USA.
AZ thanks Professor Pauchy Hwang and the Department of Physics of the
National Taiwan University, where this work was initiated,
for warm hospitality.
\\
\\
{\bf Note Added}

The Ansatz discussed in this paper has been proposed earlier by D. Black, 
A. Fariborz, S. Nasri and J. Schechter\cite{bfns}. Their theoretical 
motivation and the values of the masses they obtained are however rather 
different.

\tighten


\end{document}